\documentclass{article}

\usepackage{comment}

\usepackage{amssymb,amsfonts,amsmath}
\usepackage{cite,enumerate,float,indentfirst}
\usepackage{color}

\def\be{\begin{eqnarray}}
\def\ee{\end{eqnarray}}

\def\Tr{{\rm Tr}\,}

\definecolor{red}{rgb}{1,0,0}
\definecolor{orange}{rgb}{1,0.5,0}
\definecolor{violet}{rgb}{0.7,0,1}

\hoffset= - 1.4in         

\title{\vspace{-1cm}{\Large {\bf 
			Combinatorics of KP hierarchy structural constants
		}
		\date{}
		\author{
			{\bf A. Andreev$^{a,c}$}\thanks{andreev.av@phystech.edu},
			{\bf A. Popolitov$^{a,b,c}$}\thanks{popolit@gmail.com},
			{\bf A. Sleptsov$^{a,b,c}$}\thanks{sleptsov@itep.ru}, 
			{\bf A. Zhabin$^{a,c}$}\thanks{alexander.zhabin@yandex.ru}}
	}}

\begin{document}

\maketitle
\vspace{-4.2cm}
\hfill MIPT/TH-19/20

\hfill ITEP/TH-34/20

\hfill IITP/TH-21/20

\vspace{2.7cm}


\bigskip

\begin{center}\vspace{-1cm}
	$^a$ {\small {\it Institute for Theoretical and Experimental Physics, Moscow 117218, Russia}}\\
	$^b$ {\small {\it Institute for Information Transmission Problems, Moscow 127994, Russia}}\\
	$^c$ {\small {\it Moscow Institute of Physics and Technology, Dolgoprudny 141701, Russia }}
\end{center}

\ 

\begin{large}
\centerline{\it Dedicated to the memory of Sergey Mironovich Natanzon}
\end{large}

\ 


\bigskip

\centerline{ABSTRACT}

\bigskip

{\footnotesize
Following Natanzon-Zabrodin, we explore the Kadomtsev–Petviashvili hierarchy as an infinite system of mutually consistent relations on the second derivatives of the free energy with some universal coefficients. From this point of view, various combinatorial properties of these coefficients naturally highlight certain non-trivial properties of the KP hierarchy. Furthermore, this approach allows us to suggest several interesting directions of the KP deformation via a deformation of these coefficients. We also construct an eigenvalue matrix model, whose correlators fully describe the universal KP coefficients, which allows us to further study their properties and generalizations.
}

\bigskip

\bigskip

{ \it This paper is just the beginning of a very large program of multi-faceted study of the KP hierarchy suggested to us by Sergey Natanzon. He had his own special view of the KP hierarchy, which made it possible to see in it some new interesting structures that are completely invisible with other approaches. We are deeply grateful to him for numerous scientific discussions, for fueling our interest in the KP hierarchy and for his characteristic style of discussing science.}

\section{Introduction}
The Kadomtsev-Petviashvili (KP) hierarchy has many different applications in modern physics and mathematics. Historically it was studied as equations with soliton solutions, but very soon it was discovered that partition functions and correlators of some field theories are solutions of the hierarchy as well. It often happens that partition function can be represented as a matrix model, which provides a connection between KP hierarchy and matrix models. Probably the most famous example is the Kontsevich matrix model \cite{kontsevich1992intersection}, which is a partition function of 2D gravity. Among other important examples lattice gauge theories of QCD \cite{KHARCHEV_1995, mironov1996unitary}, the Ooguri-Vafa partition function for HOMFLY polynomials of any torus knot \cite{mironov2013character, dunin2020topological}, generating function for simple Hurwitz numbers \cite{alexandrov2012integrability, natanzon2020hurwitz, mironov2020around}. Moreover, recently interest in KP hierarchy resurgent due to fantastic rapid progress in understanding of superintegrable properties of a particular version of KP, the so-called BKP hierarchy \cite{alexandrova2020intersection, alexandrov2020kdv, natanzon2014hurwitz, mironov2020around}.

The KP hierarchy can be understood as an infinite system of compatible non-linear differential equations. All the equations may be encoded in the Hirota bilinear identity:
\begin{equation}\label{hirota_bilinear_identity}
    \oint_{\infty} e^{\xi(\overline{\textbf{t}}, z)} \, \tau(\textbf{t} + \overline{\textbf{t}} - [z^{-1}]) \, \tau(\textbf{t} - \overline{\textbf{t}} + [z^{-1}]) \, dz = 0,
\end{equation}
where we used a standard notation
\begin{equation}\label{shift_of_times}
    \begin{gathered}
    \xi(\textbf{t}, z) = \sum_{k=1}^{\infty} t_{k} z^{k}\\
    \textbf{t} \pm [z^{-1}] = \left\{ t_{1} \pm \frac{1}{z}, t_{2} \pm \frac{1}{2z^2}, t_{3} \pm \frac{1}{3z^3}, \dots \right\}
    \end{gathered}
\end{equation}
Expanding the integrand near $z=\infty$ and calculating the coefficient in the front of $z^{-1}$, each coefficient in front of every monomial of ${\bf \bar t}$ gives an equation for $\tau({\bf t})$. Functions that satisfy \eqref{hirota_bilinear_identity} are called $\tau$-functions. They may depend on an infinite number of variables $\textbf{t} = \{t_{1}, t_{2}, t_{3}, \dots \}$ called "times". Previously mentioned partition and generating functions are KP $\tau$-functions. According to the works of Kyoto school \cite{jimbo1983solitons, date1982transformation}, KP hierarchy closely related to rich mathematical structures, such as infinite-dimensional Lie algebras, projective manifolds, symmetric functions and boson-fermion correspondence. Each of these mathematical structures provides alternative language for description of KP solutions and highlights different solutions' properties. Moreover, looking at any particular solution from several points of view provides deep insights about its structure.

All mentioned examples of $\tau$-functions and many others have a geometric expansion over compact Riemann surfaces (genus expansion). Genus expansion for $\tau$-functions coincides with expansion in parameter $\hbar$ for the $\hbar$-KP hierarchy \cite{andreev2020genus}. The introduction of the $ \hbar $ parameter slightly modifies the hierarchy and allows one, among other things, to obtain solutions of the classical KP hierarchy for $ \hbar = 1 $ and dispersionless KP for $ \hbar \rightarrow 0 $ \cite{krichever1992dispersionless, dubrovin1992hamiltonian}. This $\hbar$-formulation of the KP hierarchy was first studied by Takasaki and Takebe in \cite{takasaki1992quasi, TAKASAKI_1995}, where they described a method for deformation of the classical $\tau$-function.

Natanzon and Zabrodin formulated another approach \cite{Natanzon_2016, natanzon2015symmetric} for description of the $\hbar$-KP. The advantage of their approach is that formal solutions for the $F$-function ($ F = \log \tau $) can be explicitly expressed in terms of boundary data using universal integer coefficients that help to define the entire $ \hbar $-KP hierarchy. Moreover, an arbitrary solution of the $ \hbar $-KP hierarchy can be restored from its boundary data, using these coefficients and their higher analogs, which are determined recursively. Namely, the set of the integer coefficients $P_{i,j}(s_1,\dots,s_m)$, which we also call the \textit{universal KP coefficients},
enters the KP equations as (see, for instance, \cite{dubrovin1989real})

\begin{equation}\label{equations}
    \frac{\partial^{\hbar}_i\partial^{\hbar}_j F}{ij}=\sum\limits_{m\geq 1}\frac{(-1)^{m+1}}{m}\sum\limits_{s_1,\dots,s_m \geq 1} P_{i,j}(s_1,\dots,s_m)\,\frac{\partial_{x}\partial^{\hbar}_{s_1}F}{s_1}\dots\frac{\partial_{x}\partial^{\hbar}_{s_m} F}{s_m}.
\end{equation}
where $\partial^{\hbar}_i$ is a $\hbar$-deformed derivative with respect to $t_i$, see formula \eqref{hbar_derivative} below. From these equation we see that $P_{i,j}(s_1,\dots,s_m)$ are one of the central ingredients of the KP equations. Definition of these coefficients can be given in combinatorial terms by enumeration of sequences of positive integers (see section 2, formula \eqref{P_ij_definintion}).


The main goal of this paper is to establish and develop the relation between combinatorics and integrability. We want to find out how basic properties of the combinatorial coefficients $P_{i,j}(s_1,\dots,s_m)$ affect the various properties of $\tau$-functions. The purpose of the paper is to point out new interesting research directions, but we do not develop them exhaustively in this short note.
Therefore, in many cases we stop after providing first non-trivial example, just enough to demonstrate, that a particular directions is potentially interesting and is worth studying.

The paper is organized as follows. In section 2, we introduce all the necessary definitions and theorems.

Section 3 is devoted to various approaches to calculation of the combinatorial coefficients. We show that they can be calculated using an explicit formula that includes the sum of the binomial coefficients and has a clear geometric meaning. In addition, we consider two different generating function for the universal coefficients. One of them, up to normalization, has the simple form of a sum over Young diagrams of length $\ell(\lambda)\leq 2$:
\begin{equation}\label{P_2_gf}
    F(y_1,y_2;\mathbf{x})\sim\sum\limits_{\lambda}S_{\lambda}(y_1,y_2) S_{\lambda}(\mathbf{x}),
\end{equation}
where $ S_{\lambda} $ is Schur polynomial. This generating function becomes a $\tau$-function of KP hierarchy itself after standard replacement of variables $kt_k=\sum_{i} x^k_i$, which gives us a hint on possible deformation of the universal coefficients (section 6), considering another solutions of KP hierarchy as generating function of new coefficients. 

The second generating function corresponds to the, so-called, Fay identity and, as we discuss in section 4, allows us to obtain some restrictions on resolvents in topological recursion \cite{chekhov2006free, chekhov2006hermitian, eynard2007invariants, eynard2005topological, alexandrov2004partition, alexandrov2007m, alexandrov2007instantons}.



In section 5 we construct a simple matrix eigenvalue model, whose correlators give the universal KP coefficients. The form of these correlators also makes it possible to generalize the coefficients. Generalization of matrix model has the following motivation. There are Ward identities in matrix models which can be solved recursively, and as we expect, corresponding recursion relations are related with recursion relations for higher analogs of universal coefficients in some sense. Furthermore generating function for the averages of Schur polynomials $\langle S_{\lambda_1} \dots S_{\lambda_m} \rangle$ depend on the set of time variables $\{\mathbf{t}^{(1)},\dots,\mathbf{t}^{(m)}\}$ and in the simplest case \eqref{P_2_gf} we obtain $\tau$-function of KP hierarchy, so generalized matrix model may be somehow connected with $m$-component KP hierarchy.

In Section 6 we discuss possible approach to KP deformation via deformation of generating functions of the combinatorial coefficients. We suggest another deformed generating functions that have the same properties as the initial one. Such consideration may help to understand what is the role of the combinatorial coefficients in $\hbar$-KP hierarchy: are they responsible for integrability or the certain form of equations \eqref{equations} is important.

The last section 7 is a discussion where we list main results of this paper and questions for further research.

\section{Definitions}

\textbf{Schur polynomials.} Following \cite{macdonald1998symmetric} we define Young diagram as a sequence of ordered positive integers $\lambda_1 \geq \dots \geq \lambda_{\ell(\lambda)} > 0 $ and denote it as $\lambda=[\lambda_1,\dots,\lambda_{\ell(\lambda)}]$; $\ell(\lambda)$ is the length of Young diagram. Schur polynomials $ S_{\lambda}(\mathbf{x}) $ are symmetric functions depending on an arbitrary set of variables $\mathbf{x}=\{x_1,x_2,\dots\}$ and a Young diagram $\lambda$.
\begin{equation}
    S_{\lambda}(x_1,\dots,x_n) := \frac{\det\limits_{1\leq i,j\leq n} \left(x_i^{\lambda_j+j-1}\right) }{\det\limits_{1\leq i,j\leq n} \left(x_i^{j-1}\right) }
\end{equation}
If $ n > \ell(\lambda) $, then $ \lambda_{j} $ are equal to zero for large enough $ j $. Schur polynomials labeled by Young diagrams of length $ \ell(\lambda)=1 $ we call symmetric Schur polynomials. Although all Schur polynomials are symmetric functions, such a name for particular Young diagrams is due to representation theory. Sometimes Schur polynomials are considered in variables $ \mathbf{t} = \{t_{1}, t_{2}, \dots\} $. The change from variables $ \mathbf{x} $ is given via
\begin{equation}
    t_{k} = \frac{1}{k} \sum_{i \ge 1} x_{i}^{k}.
\end{equation}
An important property of Schur polynomials that we frequently use in what follows is the Cauchy-Littlewood identity:
\begin{equation}\label{cauchy_littlewood_identity}
    \sum_{\lambda} S_{\lambda}(\textbf{t}) S_{\lambda}(\overline{\textbf{t}}) = \exp \left( \sum_{k=1}^{\infty} k t_{k} \overline{t}_{k} \right)
\end{equation}

\vspace{0.5cm}
\textbf{$\hbar$-KP hierarchy.}
We briefly review the main facts about the KP equations and solutions. For the detailed explanation see \cite{miwa2000solitons}. KP hierarchy is an infinite set of non-linear differential equations with the first equation given by
\begin{equation}
	\frac{1}{4} \frac{\partial^{2} F}{\partial t_{2}^{2}} = \frac{1}{3} \frac{\partial^{2} F}{\partial t_{1} \partial t_{3}} - \frac{1}{2} \left( \frac{\partial^{2} F}{\partial t_{1}^{2}} \right)^{2} - \frac{1}{12} \frac{\partial^{4} F}{\partial t_{1}^{4}}
\end{equation}
It is more common to work with $ \tau $-function $ \tau(\textbf{t}) = \exp(F(\textbf{t})) $ than with free energy $ F(\textbf{t}) $. We assume that $ \tau(\textbf{t}) $ is at least a formal power series in times $ t_{k} $, and maybe it is even a convergent series. Entire set of equations of hierarchy can be written in terms of $ \tau $-function using Hirota bilinear identity \eqref{hirota_bilinear_identity}, which, in turn, is equivalent to the following functional equation
\begin{equation}\label{Hirota_functional_relation}
	(z_{1} - z_{2})\tau^{[z_{1},z_{2}]}\tau^{[z_{3}]} + (z_{2} - z_{3})\tau^{[z_{2},z_{3}]}\tau^{[z_{1}]} + (z_{3} - z_{1})\tau^{[z_{3},z_{1}]}\tau^{[z_{2}]} = 0
\end{equation}
where
\begin{equation}\label{time_shift_classical}
	\tau^{[z_{1}, \dots, z_{m}]}(\textbf{t}) = \tau \left( \textbf{t} + \sum_{i=1}^{m}[z_{i}^{-1}] \right)\\
\end{equation}
and the shift of times is the same as in \eqref{shift_of_times}. Equation \eqref{Hirota_functional_relation} should be satisfied for an arbitrary $ z_{1}, z_{2}, z_{3} $. One can expand $ \tau $-function at the vicinity of $ z_{i} = \infty $ and obtain partial differential equation for $ \tau $-function at every term $ z_{1}^{-k_{1}} z_{2}^{-k_{2}} z_{3}^{-k_{3}} $.
	
All formal power series solutions of KP hierarchy can be decomposed over the basis of Schur polynomials
\begin{equation}\label{bosonic_representation}
	\tau(\textbf{t}) = \sum_{\lambda} C_{\lambda} S_{\lambda}(\textbf{t}).
\end{equation}
Function written as a formal sum over Schur polynomials is a KP solution if and only if coefficients $ C_{\lambda} $ satisfy the Pl\"{u}cker relations. The first such relation is
\begin{equation}\label{plucker_relations}
	C_{[2,2]} C_{[\varnothing]} - C_{[2,1]} C_{[1]} + C_{[2]} C_{[1,1]} = 0.
\end{equation}

The simplest way to define $\hbar$-KP hierarchy is to deform bilinear equations \eqref{Hirota_functional_relation} for $\tau$-function of the classical KP hierarchy in the following way \cite{Natanzon_2016, dubrovin1989real}:
\begin{equation}\label{h_KP_equation}
	\begin{gathered}
		(z_{1} - z_{2})\tau^{[z_{1},z_{2}]}\tau^{[z_{3}]} + (z_{2} - z_{3})\tau^{[z_{2},z_{3}]}\tau^{[z_{1}]} + (z_{3} - z_{1})\tau^{[z_{3},z_{1}]}\tau^{[z_{2}]} = 0\\
		\tau^{[z_{1}, \dots, z_{m}]}(\textbf{t}) = \tau \left( \textbf{t} + \hbar \sum_{i=1}^{m}[z_{i}^{-1}] \right)\\
		\textbf{t} + \hbar [z^{-1}] = \left\{ t_{1} + \frac{\hbar}{z}, t_{2} + \frac{\hbar}{2z^{2}}, t_{3} + \frac{\hbar}{3z^{3}}, \dots \right\}
	\end{gathered}
\end{equation}
By setting parameter $ \hbar = 1 $ we obtain classical KP hierarchy and the limit $ \hbar \rightarrow 0 $ provides celebrated dispersionless hierarchy \cite{krichever1992dispersionless, dubrovin1992hamiltonian}. The other equivalent way to encode all the ($\hbar$-)KP equations is the differential Fay identity:
\begin{equation}\label{fay_identity}
    \Delta(z_1)\Delta(z_2)F= \log \left(1-\frac{\Delta(z_1)\partial_{1}F-\Delta(z_2)\partial_{1}F}{z_1-z_2}\right),
\end{equation}
where
\begin{equation}
    \Delta(z)=\frac{e^{\hbar D(z)}-1}{\hbar}, \:\:\: D(z)=\sum\limits_{k\geq 1}\frac{z^{-k}}{k}\partial_{k}.
\end{equation}

KP hierarchy can be considered as an infinite set of compatible differential equations on the $ F $-function, where $ F(\textbf{t})=\hbar^2\log(\tau(\textbf{t})) $. To describe the equations in an unfolded form we need two more definitions. First one is deformed partial derivatives $ \partial_{k}^{\hbar} $ which are defined via symmetric Schur polynomials in $ \textbf{t} $-variables. Each $ t_{i} $ one should replace with $ \frac{\hbar}{i}\partial_{i} $:
\begin{equation}\label{hbar_derivative}
	\partial_{k}^{\hbar} := \frac{k}{\hbar} S_{[k]}(\hbar \widetilde{\partial}), \;\;\;\;\; \widetilde{\partial} = \left\{ \partial_{1}, \frac{1}{2} \partial_{2}, \frac{1}{3} \partial_{3}, \dots \right\}
\end{equation}
Limit $ \hbar \rightarrow 0 $ transforms deformed derivatives $ \partial_{k}^{\hbar} $ into usual ones $ \partial_{k} $.

The next definition is the main topic of our study. Let us define combinatorial coefficients $ P_{i,j}(s_{1}, \dots, s_{m}) $ as the number of sequences $(i_1,\dots,i_m)$ and $(j_1,\dots,j_m)$ of positive integers such that $i_1+\dots +i_m=i$, $j_1+\dots+j_m=j$ and $i_k+j_k=s_k+1$. These coefficients can also be understood as the number of matrices of size $ 2 \times m $ with fixed sums over rows and columns:
\begin{equation}\label{P_ij_definintion}
    \boxed{P_{i,j}(s_1,\dots,s_m):=\# \left\{
    \begin{pmatrix} 
        i_{1} & \dots & i_{m} \\
        j_{1} & \dots & j_{m} \\
    \end{pmatrix} \Bigg|
    i_k,j_k\in \mathbb{N},
    \begin{array}{cc}
        i_1+\dots+i_m=i \\
        j_1+\dots+j_m=j \\
        i_k+j_k=s_k+1 \;\; \forall k \in \overline{1, m}
    \end{array}
    \right\} }
\end{equation}
Coefficients \eqref{P_ij_definintion} are fundamental in the following sense. They allow us to express all the KP equations in an explicit form and fully determine $\hbar$-KP hierarchy.

Following \cite[Lemma 3.2]{Natanzon_2016} the $\hbar$-KP hierarchy can be rewritten as the system of equations:
\begin{equation}\label{NZ_formulation}
    \frac{\partial^{\hbar}_i\partial^{\hbar}_j F}{ij}=\sum\limits_{m\geq 1}\frac{(-1)^{m+1}}{m}\sum\limits_{s_1,\dots,s_m \geq 1} P_{i,j}(s_1,\dots,s_m)\frac{\partial_{x}\partial^{\hbar}_{s_1}F}{s_1}\dots\frac{\partial_{x}\partial^{\hbar}_{s_m} F}{s_m}
\end{equation}
for the function $F(x;\mathbf{t})=F(t_1+x,t_2,t_3,\dots)$. Note that sum in the r.h.s. of  \eqref{NZ_formulation} is finite. For fixed $ i $ and $ j $ there is a restriction on $ s_{k} $. Sum of all matrix elements is a sum of rows which should coincide with a sum of columns: $ i + j = s_{1} + \dots + s_{m} + m $. For large enough values of $ s_{k} $ or a large number $ m $ coefficients $ P_{i,j}(s_{1}, \dots, s_{m}) $ are equal to zero.

\vspace{0.5cm}
The next step is to determine all the solutions of the hierarchy. For this reason we need Cauchy-like data, which is a set of functions of variable $ x $: $ \partial_{k}^{\hbar} F^{\hbar}(x,\textbf{t}) \lvert_{\textbf{t}=0} = f_{k}^{\hbar}(x) $. If we consider formal solutions, i.e. not necessarily converging series, any solution can be expressed through Cauchy-like data using universal coefficients $P^{\hbar}_{\lambda}\begin{pmatrix} s_{1} \dots s_{m} \\ l_{1} \dots l_{m} \end{pmatrix}$, which were mentioned before as higher analogs of coefficients $P_{i,j}(s_{1}, \dots, s_{m})$.

It was shown by Natanzon and Zabrodin \cite[Theorem 4.3]{Natanzon_2016} that for an arbitrary set of smooth functions
	$$ \textbf{f} = \{f_{0}^{\hbar}(x), f_{1}^{\hbar}(x), \dots\} $$
		there exists a unique solution $ F^{\hbar}(x,\textbf{t}) $ of the $\hbar$-KP hierarchy with Cauchy-like data $\textbf{f}$. This solution is of the form
		\begin{equation}\label{F_function_solution}
			F^{\hbar}(x,\textbf{t}) = f_{0}^{\hbar}(x) + \sum_{|\lambda| \ge 1} \frac{f_{\lambda}^{\hbar}(x)}{\sigma(\lambda)} t_{\lambda}^{\hbar}
		\end{equation}
		where $ f_{[k]}^{\hbar}(x) = f_{k}^{\hbar}(x) $ and 
		\begin{equation}\label{f_coefficients}
			f_{\lambda}^{\hbar}(x) = \sum_{m \ge 1} \sum_{\substack{s_{1} + l_{1} + \dots + s_{m} + l_{m} = |\lambda|\\ 1 \le s_{i}; \; 1 \le l_{i} \le l(\lambda) - 1}} P^{\hbar}_{\lambda}
			\begin{pmatrix}
				s_{1} \dots s_{m} \\
				l_{1} \dots l_{m}
			\end{pmatrix}
			\partial_{x}^{l_{1}} f_{s_{1}}^{\hbar}(x) \dots \partial_{x}^{l_{m}} f_{s_{m}}^{\hbar}(x)
		\end{equation}
		for $ l(\lambda) > 1 $. $ \sigma(\lambda) = \prod_{i\ge1}m_{i}! $, where exactly $ m_{i} $ parts of the partition $\lambda$ have length $ i $.

The full recursive definition of universal coefficients $ P_{\lambda}^{\hbar} $ is quite unwieldy and can be found in \cite{Natanzon_2016}. In this paper we are interested in simplest coefficients with $ l_1 = \dots = l_m = 1 $ and $\lambda=[i,j]$. They are defined as coefficients \eqref{P_ij_definintion} with the normalization factor
\begin{equation}
   P^{\hbar}_{[i,j]}\begin{pmatrix}  s_{1} & \dots & s_{m} \\ 1 & \dots & 1 \end{pmatrix}:=\frac{(-1)^{m+1}ij}{m \cdot s_1\dots s_m}P_{i,j}(s_1,\dots,s_m)
\end{equation}
The other coefficients with $\ell(\lambda)\geq 2$ and $l_i>1$ can be obtained from \eqref{P_ij_definintion} using recursion relations.

\section{Remarkable properties of combinatorial coefficients $ P_{i,j}(s_1, \dots, s_m) $}
As it was claimed \eqref{NZ_formulation}, we can rewrite all KP equations with help of certain combinatorial coefficients $ P_{i,j}(s_1, \dots, s_m) $. So it is natural to ask if there is some connection between properties of KP hierarchy and properties of these combinatorial objects.
Therefore, in this section we recall the most prominent properties
of the constants $ P_{i,j}(s_1, \dots, s_m) $, as well as the context around their combinatorics.
We postpone the discussion of the connection with the KP till the next section.

Coefficients $ P_{i,j}(s_1, \dots, s_m) $ and their n-point generalizations \eqref{number_of_matrices},
in fact, arise in the theory of flow networks \cite{barvinok2016combinatorics} and are very well studied. Standard problem in the theory of flow networks is finding the maximum flow which gives the largest total flow from the source to the sink. We interested here in more simple question: what is the number of different flows on the graph where all $n$ sources and $m$ sinks are connected by edges which is exactly coefficients $P_{i_1,\dots,i_n}(s_1,\dots,s_m)$.

Since there is a rich combinatorial structure of the combinatorial coefficients,
there are many different ways to calculate them, each having potential implications
for our topic: explicit formula as the sum over vertices of hypercube, recursion formula and generating function.

\begin{itemize}
    \item First of all, there is an explicit approach to calculation of the coefficients using geometric interpretation and inclusion-exclusion principle: combinatorial coefficients $P_{i,j}(s_1,\dots,s_m)$ can be represented as the sum over vertices of $m$-dimensional hypercube
    \begin{equation}\label{P_formula}
        P_{i,j}(s_1,\dots,s_m)=\delta_{s_1+\dots+s_m+m,i+j}\sum\limits_{\{\sigma_k=\{0,1\}|k=1,\dots,m\}}(-1)^{\sigma_1+\dots+\sigma_m}{i-\sigma_1 s_1-\dots-\sigma_m s_m-1\choose m-1}
    \end{equation}
    The cube is parametrized by the sequences of zeros and unities $(\sigma_1,\dots,\sigma_m)$.
    Note here that we take binomial coefficients ${m\choose k}$ equal to zero if $m<k$ or $m<0$ or $k<0$. (see Appendix A for the details on the derivation)
    
    \item There is a natural generalization of the combinatorial coefficients in the following way. Matrices of size $ 2 \times m $ are distinguished in KP theory, but from the point of view of combinatorics one may consider the number of matrices of size $ n \times m $ with fixed sums over rows and columns.
    \begin{equation}\label{number_of_matrices}
        P_{i_1\dots i_n}(s_1,\dots,s_m) := \# \left\{
        \begin{pmatrix} 
            i_{1}^{(1)} & \dots & i_{m}^{(1)} \\
            \vdots & \ddots & \vdots \\
            i_{1}^{(n)} & \dots & i_{m}^{(n)} \\
        \end{pmatrix} \Bigg| 
    i_{k}^{(l)} \in \mathbb{N}, \;
    \begin{array}{cc}
        i_{1}^{(l)} + \dots + i_{m}^{(l)} = i_{1}, & \forall l \in \overline{1,n} \\
        i_{k}^{(1)} + \dots + i_{k}^{(n)} = s_{k} + n - 1, & \forall k \in \overline{1, m} 
    \end{array}
    \right\}
    \end{equation}
    
    Such objects arise in the simplest flow network problem: it is the number of integer flows on complete bipartite graph \cite{barvinok2016combinatorics}.
    
    Note that defined coefficients are symmetric up to permutation within the set of parameters $ s_{k} $ and within the set of indices $ i_{l} $. Thus, one may consider an ordered sets of indices and parameters labeled by Young diagrams $ \lambda $ and $ \mu $. More information about  combinatorial meaning and different applications of such coefficients, denoted as $ N(\lambda, \mu) $, can be found in \cite{stanley1999enumerative}. 
    
    This interpretation in terms of the number of certain matrices \eqref{number_of_matrices} allows one to obtain the following recursion relations \cite{natanzon1992differential}:
    
    \begin{equation}\label{I_recursion}
        P_{i_1\dots i_n}(s_1,\dots,s_m)=
    \sum\limits_{\left\{{i_n^1+\dots+i_n^m=i_n \atop 1\leq i_n^l\leq s_l|l=1,\dots,m}\right\}}P_{i_1\dots i_{n-1}}(s_1-i_1^n+1,\dots,s_m-i_m^n+1)
    \end{equation}
    and
    \begin{equation}\label{S_recursion}
        P_{i_1\dots i_n}(s_1,\dots,s_m)=
    \sum\limits_{\left\{{i_1^m+\dots+i_n^m=s_m+n-1 \atop 1\leq i_l^m\leq i_l-m+1|l=1,\dots,n}\right\}}P_{i_1-i_1^m,\dots, i_{n}-i_{n}^m}(s_1,\dots,s_{m-1})
    \end{equation}
    Note that \eqref{I_recursion} and \eqref{S_recursion} are the same up to the symmetry between indices $\{i_l\}$ and parameters $\{s_l\}$ mentioned above.
    
    \item The last approach to calculation of the combinatorial coefficients is by means of generating function. We can construct such function in two different ways. Both highlight some interesting properties on the KP hierarchy side which we discuss in  section 4.  
    Firstly, we can write it in the following way:
\begin{equation}
    \widetilde{G}_{nm}(\mathbf{x},\mathbf{y})=\sum\limits_{i_1\geq 1,\dots,i_n\geq 1}y_1^{i_1}\dots y_n^{i_n}\sum\limits_{s_1\geq 1,\dots, s_m\geq 1}x_1^{s_1}\dots x_m^{s_m}P_{i_1\dots i_n}(s_1,\dots,s_m)=\left(\prod\limits_{l=1}^m x_l\right)\left(\prod\limits_{k=1}^n  y_k^m\right) \sum\limits_{\lambda} S_{\lambda}(\mathbf{x}) S_{\lambda}(\mathbf{y})
\end{equation}

    which can be rewritten more naturally with the help of shifts: $i_k^1+\dots+i_k^m=i_k\mathbf{+m}$ for $k=1,\dots,n$ and $i_1^l+\dots+i_n^l=s_l\mathbf{+n}$ for $l=1,\dots,m$:
  
    \begin{equation}\label{P_generating_function}
        G_{mn}(\mathbf{x},\mathbf{y}) = \sum\limits_{\lambda} S_{\lambda}(\mathbf{x}) S_{\lambda}(\mathbf{y}) = \prod\limits_{i,j}\frac{1}{1 - x_i y_j}
    \end{equation}
    This formula is well known \cite{stanley1999enumerative}, but we give a short calculation in Appendix B that shows how it follows from recursion relations \eqref{I_recursion}.
    
    We also consider another generating function in variables $p_k$:
    \begin{equation}\label{H_2}
        H(\mathbf{p};y_1,y_2)=\sum\limits_{m\geq 0}\frac{(-1)^{m+1}}{m}\sum\limits_{ij}y_1^i y_2^j\sum\limits p_{s_1}\dots p_{s_m}P_{ij}(s_1,\dots,s_m)=\ln\left(1+y_1y_2\sum\limits_{k=1}^{\infty}p_k\frac{y_1^k-y_2^k}{y_1-y_2}\right)
    \end{equation}
    The choice of these variables is motivated by the formula \eqref{NZ_formulation} where factors $\partial_{x}\partial_{s}F$ are included in the equation in the same way as $p_i$ into this generating function. The formula \eqref{H_2} can be obtained from the first generating function \eqref{P_generating_function} by replacement $p_k=\sum_{i}x_i^k$ (more detailed calculation can be found in Appendix B)
\end{itemize}

\section{Connection with KP hierarchy}
Now we discuss, what does the explicit form of the generating functions \eqref{P_generating_function},\eqref{H_2} mean for the KP hierarchy.
First of all, we argue that the generating function \eqref{P_generating_function} becomes the KP tau-function after some simple change of variables, which will become effective in section 6.2 where we describe possible deformations.

Second of all, the other generating function \eqref{H_2} allows one to easily derive Fay-identity form
of the KP hierarchy.

We also discuss here interpretation of the combinatorial formula \eqref{f_coefficients} in terms of solutions that can be restored using topological recursion.

\begin{itemize}
    \item  Generating function \eqref{P_generating_function} of the redefined coefficients can be rewritten in another variables by replacement $kt_k=\sum\limits_{i}x_i^{k}$ and $k\bar{t}_k=\sum\limits_{i}y_i^{k}$. In these variables, using Cauchy-Littlewood identity \eqref{cauchy_littlewood_identity} we obtain:
    \begin{equation}
        G(\mathbf{x},\mathbf{y})=\sum_{\lambda}S_{\lambda}(\mathbf{t})S_{\lambda}(\mathbf{\bar{t}})=e^{\sum_{k}k t_k\bar{t}_k}
    \end{equation}
    which is trivially a $\tau$-function of KP (or Toda) hierarchy where $\mathbf{t}$ and $\mathbf{\bar{t}}$ are corresponding times. So the generating function for coefficients, which defines $\hbar$-KP, is the trivial $\tau$-function itself. We will discuss this property in section 6 trying to deform the combinatorial coefficients. 
    
    \item The second generating function allows us to write the analog of the Fay identity in the following way: it gives us generating function for all KP equations \eqref{NZ_formulation} by replacement $p_k\rightarrow \frac{\partial\partial_{k}^{\hbar}F}{k}$:
    
    \begin{equation}\label{Fay_identity}
         \frac{\partial^{\hbar}_i\partial^{\hbar}_j F}{ij}=\left[y_1^i y_2^j\right]\ln\left(1+y_1y_2\sum\limits_{k=1}^{\infty}\frac{\partial\partial^{\hbar}_k F}{k}\frac{y_1^k-y_2^k}{y_1-y_2}\right)
    \end{equation}
    
    From the other hand the Fay identity for $\hbar$-KP hierarchy has the form \eqref{fay_identity}. Now, using replacement $z_i\rightarrow \frac{1}{y_i}$ and the fact that $\partial_1=\partial_{x}=\partial$ we obtain exactly \eqref{Fay_identity}. The explicit derivation of \eqref{Fay_identity} from Fay identity can be found in \cite{Natanzon_2016}.
    
    As we can see here, combinatorial properties of the coefficients $P_{i,j}(s_1,\dots,s_m)$ in the form \eqref{I_recursion} lead to the generating function \eqref{H_2} which is exactly Fay identity in terms of KP hierarchy.
    
    \item Let us now turn to the question of restrictions which explicit form of KP equations imposes on the topological recursion. Many solutions of the KP hierarchy (e.g., simple Hurwitz numbers \cite{alexandrov2012integrability}, Hermitian matrix model \cite{gerasimov1991matrix, kharchev1991matrix, kharchev1993generalized}, Kontsevich $\tau$-function \cite{kontsevich1992intersection}) allows one to construct multi-differentials, which are related by the so-called spectral curve topological recursion \cite{chekhov2006free, chekhov2006hermitian, eynard2007invariants, eynard2005topological, alexandrov2004partition, alexandrov2007m, alexandrov2007instantons}. The initial data for the recursion procedure are 1-point and 2-point function of genus 0 which are expected to be independent. However, naively, from formula \eqref{f_coefficients} it follows that two-point functions $f^{\hbar}_{\lambda_1,\lambda_2}$ can be expressed via one-point functions $f^{\hbar}_{k}$. 
    
    Let us recall main concepts of the topological recursion. This approach firstly arose in the theory of matrix models where all correlators have natural genus expansion \cite{Mironov_2017, Eynard_2004}. In such theories we consider the following correlators which are called resolvents:
    \begin{equation}
        W_n(p_1,\dots,p_n)=\left\langle \Tr \frac{1}{p_1-X} \dots \Tr \frac{1}{p_n-X} \right\rangle_{Conn}
    \end{equation}
    where we integrate over matrices $X$ with some measure and $Conn$ means we consider connected diagrams only. They also have some genus expansion
    \begin{equation}
        W_n=\sum\limits_{g}\hbar^{2g} W_{g,n}
    \end{equation}
    Topological recursion allows us to recover all resolvents in the genus $g=n$ from $g<n$ resolvents if we know the initial data: spectral curve, $W_{0,1}$ and $W_{0,2}$.
    
    Moreover, in many cases where topological recursion is applicable, the logarithm of partition function $F=\hbar^2\log(Z)$ turns out to be a solution of $\hbar$-KP. We can also represent resolvents via $F$ in the following way
    \begin{equation}
        W(p_1,\dots,p_s)=-\frac{\partial}{\partial V(p_1)}\dots \frac{\partial}{\partial V(p_s)}F\Big|_{\mathbf{t}=0,x=0}
    \end{equation}
    where
    \begin{equation}
        \frac{\partial}{\partial V(p)}=-\sum\limits_{j=1}^{\infty}\frac{1}{p^{j+1}}\frac{\partial}{\partial t_{j}}
    \end{equation}
    is the loop insertion operator \cite{alexandrov2004partition}.
    
    Returning to the Natanzon-Zabrodin formulation of KP hierarchy we can consider the Cauchy-like data as genus zero resolvents since in the limit $\hbar\rightarrow 0$ formula \eqref{F_function_solution} gives:
    \begin{equation}
		F^{\hbar=0}(x,\textbf{t}) = f_{0}^{\hbar=0}(x) + \sum_{|\lambda| \ge 1} \frac{f_{\lambda}^{\hbar=0}(x)}{\sigma(\lambda)} t_{\lambda}
	\end{equation}
    and
    
    \begin{equation}
        W_0(p_1,\dots,p_n)=(-1)^n\sum\limits_{\lambda_1, \dots, \lambda_n\geq 1}\frac{1}{p_1^{\lambda_1+1}}\dots\frac{1}{p_n^{\lambda_n+1}}\partial_1\dots \partial_n F\Big|_{\mathbf{t}=0,x=0,\hbar=0}=(-1)^n\sum\limits_{\lambda_1\geq \dots\geq \lambda_n\geq 1}\frac{1}{p_1^{\lambda_1+1}}\dots\frac{1}{p_n^{\lambda_n+1}} f^{\hbar=0}_{\lambda}\Big|_{x=0}
    \end{equation}
    
    Now it is clear that KP hierarchy imposes some restrictions since this formula connects two point resolvents with functions $\partial f_{k}|_{x=0}$ which in terms of $W$ corresponds to two-point resolvents in the following way. Let 
    \begin{equation}
        W_0(p_1,p_2)=\sum\limits_{\lambda_1\geq\lambda_2\geq1}\frac{1}{p_1^{\lambda_1+1}p_2^{\lambda_2+1}}\omega_{\lambda_1\lambda_2}
    \end{equation}
    then $f^{\hbar=0}_{\lambda_1\lambda_2}=\omega_{\lambda_1\lambda_2}$ and $\omega_{\lambda_1 1}=\partial f^{\hbar=0}_{\lambda_1}|_{x=0}$. It is possible now to write nontrivial condition on two-point resolvents using \eqref{f_coefficients} for $\ell(\lambda)=2$:
    \begin{equation}
        \boxed{\frac{\omega_{\lambda_1,\lambda_2}}{\lambda_1\lambda_2}=\left[y_1^{\lambda_1}y_2^{\lambda_2}\right]\ln\left(1+y_1y_2\sum\limits_{k=1}^{\infty}\frac{\omega_{k,1}}{k}\frac{y_1^k-y_2^k}{y_1-y_2}\right)}
    \end{equation}
    Summarizing, the combinatorial view on KP hierarchy allows us to obtain a nontrivial condition on solutions of KP hierarchy that admits recovering using topological recursion. This equation means that we can express all genus zero two-point resolvents using only $\omega_{k,1}$ data. It would be very interesting to see whether these KP restrictions are related with the decomposition property (\cite[Lemma 4.1]{eynard2011invariants}), which under certain mild assumptions holds for $W_{0,2}$. This question is left for further research.
    
\end{itemize}

\section{Eigenvalue model}
In this section we provide a complete description of combinatorial coefficients $ P_{i_{1}, \dots, i_{n}}(s_{1}, \dots, s_{m}) $ in terms of an eigenvalue model. The model is an integral over eigenvalues of a matrix with a simple measure. Combinatorial coefficients appear to be certain correlators in the model, i.e. averages of product of $m$ symmetric Schur polynomials $ \langle S_{s_{1}-1} \dots S_{s_{m}-1} \rangle $. An arbitrary correlator in the model may be expressed with the help of the full basis of observables. The basis is obtained as a natural generalization of combinatorial coefficients $ P_{i_{1}, \dots, i_{n}}(s) $ with one parameter $s$ and coincides with a subset of Kostka numbers. Partition function of the model can be calculated explicitly. The common property of matrix models is the existence of Ward identities that might be solved recursively. In this model Ward identities give new recursion relations on combinatorial coefficients $ P_{i_{1}, \dots, i_{n}}(s_{1}, \dots, s_{m}) $.

This model takes the simplest form for slightly modified coefficients $ P_{i_{1}, \dots, i_{n}}(s_{1}, \dots, s_{m}) $ with symmetric definition for both lower indices $ i_{k} $ and integers $ s_{j} $:
\begin{equation}\label{definition_shifted}
    \begin{gathered}
        i_{1}^{(k)} + \dots + i_{m}^{(k)} = i_{k} + m - 1, \;\;\; 1 \le k \le n\\
        i_{j}^{(1)} + \dots + i_{j}^{(n)} = s_{j} + n - 1, \;\;\; 1 \le j \le m
    \end{gathered}
\end{equation}
Note that such a definition differs from \eqref{number_of_matrices} by shift of $i_{k}$. However, both definitions provide coefficients that are in one-to-one correspondence by the shift of indices, so we denote them as $ P_{i_{1}, \dots, i_{n}}(s_{1}, \dots, s_{m}) $ as well.

Let us introduce an eigenvalue model
\begin{equation}\label{eigenvalue_model}
    \mathcal{Z}_{n}(\textbf{t}) = \frac{1}{(2\pi i)^{n}} \oint dz_{1} \dots \oint dz_{n} \left( \prod_{k=1}^{n} z_{k}^{-i_{k}} \right) \exp \left( \sum_{k=1}^{\infty} t_{k} \Tr Z^{k} \right),
\end{equation}
where $ z_{k} $ are complex variables, integration contours are unit circles and $ Z $ is diagonal matrix $ Z = \text{diag}(z_{1}, \dots, z_{n}) $. Using Cauchy-Littlewood identity \eqref{cauchy_littlewood_identity}, we rewrite it in the form
\begin{equation}\label{eigenvalue_Schur}
    \mathcal{Z}_{n}(\textbf{t}) = \sum_{\lambda} \left\{ \frac{1}{(2\pi i)^{n}} \oint dz_{1} \dots \oint dz_{n} \left( \prod_{k=1}^{n} z_{k}^{-i_{k}} \right) S_{\lambda}(Z) \right\} S_{\lambda}(\textbf{t}) \equiv \sum_{\lambda} \langle S_{\lambda} \rangle S_{\lambda}(\textbf{t}),
\end{equation}
which can be understood as a generating function for correlators $ \langle S_{\lambda} \rangle $. Combinatorial coefficients $ P_{i_{1}, \dots, i_{n}}(s_{1}, \dots, s_{m}) $ appear to be correlators of specific form in such eigenvalue model.

Any combinatorial coefficient $ P_{i_{1}, \dots, i_{n}}(s_{1}, \dots, s_{m}) $ can be represented as an average of $ m $ symmetric Schur polynomials:
\begin{equation}\label{coefficients_as_average}
    P_{i_{1}, \dots, i_{n}}(s_{1}, \dots, s_{m}) = \frac{1}{(2 \pi i)^{n}} \oint dz_{1} \dots \oint dz_{n} \left( \prod_{k=1}^{n} z_{k}^{-i_{k}} \right) \left( \prod_{j=1}^{m} S_{s_{j}-1}(Z) \right) \equiv \langle S_{s_{1}-1} \dots S_{s_{m}-1} \rangle
\end{equation}
Although this integral seems complicated, it i fact, has simple meaning of extracting certain coefficient in front of certain powers of $z$-variables of integrand: $ P_{i_{1}, \dots, i_{n}}(s_{1}, \dots, s_{m}) = [z_{1}^{i_{1}-1} \dots z_{n}^{i_{n}-1}]\left( \prod_{j=1}^{m} S_{s_{j}-1}(Z) \right) $. This formula can be obtained as follows. Restrictions \eqref{definition_shifted} allow us to represent the definition of combinatorial coefficients as a sum over product of delta-symbols (each restriction corresponds to one delta-symbol):
\begin{equation}
    P_{i_{1}, \dots, i_{n}}(s_{1}, \dots, s_{m}) = \sum_{\substack{i_{1}^{(1)} \ge 1\\ \dots \\i_{m}^{(1)} \ge 1}} \dots \sum_{\substack{i_{1}^{(n)} \ge 1\\ \dots \\i_{m}^{(n)} \ge 1}} \left( \prod_{k=1}^{n} \delta_{i_{1}^{(1)} + \dots + i_{m}^{(1)}, i_{k} + m - 1} \right) \left( \prod_{j=1}^{m} \delta_{i_{j}^{(1)} + \dots + i_{j}^{(n)}, s_{j} + n - 1} \right)
\end{equation}
Delta-symbols are replaced with contour integrals with the help of simple relation
\begin{equation}
    \delta_{n,m} = \frac{1}{2 \pi i} \oint dz z^{n-m-1}.
\end{equation}
We change the first $n$ delta-symbols to integrals in such way. The obtained expression is of the form 
\begin{equation}\label{almost_average_of_Schurs}
    P_{i_{1}, \dots, i_{n}}(s_{1}, \dots, s_{m}) = \frac{1}{(2 \pi i)^{n}} \oint dz_{1} \dots \oint dz_{n} \left( \prod_{k=1}^{n} z_{k}^{-i_{k}} \right) \prod_{j=1}^{m} \left[ \sum_{i_{j}^{(1)} \ge 1} \dots \sum_{i_{j}^{(n)} \ge 1} z_{1}^{i_{j}^{(1)}-1} \dots z_{n}^{i_{j}^{(n)}-1} \delta_{i_{j}^{(1)} + \dots + i_{j}^{(n)}, s_{j} + n - 1} \right]
\end{equation}
The expression in square brackets can be evaluated independently for each $j$ and depends only on $ s_{j} $. It is equal to Schur polynomial $ S_{s_{j}-1}(z_{1}, \dots, z_{n}) $. Detailed calculations are presented in Appendix C. Thus, we proved formula \eqref{coefficients_as_average}.

Eigenvalue model \eqref{eigenvalue_Schur} is a natural generalization of coefficients $ P_{i_{1}, \dots, i_{n}}(s) = \langle S_{s-1} \rangle $, i.e. one may consider averages of an arbitrary Schur polynomial $ \langle S_{\lambda} \rangle $, not only symmetric ones. Any other coefficients such as $ P_{i_{1}, \dots, i_{n}}(s_{1}, \dots, s_{m}) = \langle S_{s_{1}-1} \dots S_{s_{m}-1} \rangle $ or their natural generalizations $ \langle S_{\lambda_{1}} \dots S_{\lambda_{m}} \rangle $ can be expressed in terms of linear combinations of $ \langle S_{\lambda} \rangle $: product of Schur polynomials is decomposed in linear combination of single Schur polynomials with Littlewood-Richardson coefficients \cite{macdonald1998symmetric}. So, correlators $ \langle S_{\lambda} \rangle $ form the appropriate full basis in the space of observables of the model.

Moreover, correlators $ \langle S_{\lambda} \rangle $ coincide with Kostka numbers. One of the definitions of Kostka numbers $ K_{\lambda, \mu}$ is the decomposition of Schur polynomial into the sum over monomial symmetric functions $ m_{\lambda}(z_{1}, \dots, z_{n}) $ or, equivalently, into the sum over all weak compositions $\alpha$ of $n$ \cite{stanley1999enumerative}:
\begin{equation}
    S_{\lambda}(z_{1}, \dots, z_{n}) = \sum_{\mu} K_{\lambda, \mu} m_{\mu}(z_{1}, \dots, z_{n}) = \sum_{\alpha} K_{\lambda, \alpha} z^{\alpha},
\end{equation}
where $ z^{\alpha} $ denotes the monomial $ z_{1}^{\alpha_{1}} \dots z_{n}^{\alpha_{n}} $. The simple form of average \eqref{eigenvalue_Schur} exactly coincides with coefficient in front of one monomial in Schur polynomial decomposition: $ \langle S_{\lambda}(Z) \rangle = [z_{1}^{i_{1}-1} \dots z_{n}^{i_{n}-1}]S_{\lambda}(Z) $. The latter one is the Kostka number $ K_{\lambda, \widetilde{\alpha}} $, where $ \widetilde{\alpha} = (i_{1}-1, \dots, i_{n}-1) $. Finally, we can write
\begin{equation}\label{kostka_numbers}
    \langle S_{\lambda}(Z) \rangle = K_{\lambda, \widetilde{\alpha}}.
\end{equation}
The set of basis observables in the eigenvalue model is a subset of Kostka numbers. All correlators in the model may be expressed with the help of Kostka numbers.

\vspace{0.5cm}
The complete information about eigenvalue model is given by an explicit expression for generating function \eqref{eigenvalue_Schur}. It is possible to calculate not only $ \mathcal{Z}_{n}(\textbf{t}) $ but also more general generating function:
\begin{equation}
    \mathcal{Z}_{n}(\textbf{t}^{(1)}, \dots, \textbf{t}^{(m)}) = \sum_{\lambda_{1}} \dots \sum_{\lambda_{m}} \langle S_{\lambda_{1}} \dots S_{\lambda_{m}} \rangle S_{\lambda_{1}}(\textbf{t}^{(1)}) \dots S_{\lambda_{m}}(\textbf{t}^{(m)}),
\end{equation}
where $ \textbf{t}^{(k)} $ is an infinite vector of times $ \textbf{t}^{(k)} = (t_{1}^{(k)}, t_{2}^{(k)}, t_{3}^{(k)}, \dots ) $ for each $ k $. First of all, Cauchy-Littlewood identity \eqref{cauchy_littlewood_identity} allows us to evaluate sums over partitions $ \lambda_{1}, \dots, \lambda_{m} $ and obtain an expression similar to \eqref{eigenvalue_model}:
\begin{equation}
    \mathcal{Z}_{n}(\textbf{t}^{(1)}, \dots, \textbf{t}^{(m)}) = \frac{1}{(2\pi i)^{n}}\oint dz_{1} \dots \oint dz_{n} \left( \prod_{k=1}^{n} z_{k}^{-i_{k}} \right) \left( \prod_{k=1}^{m} \exp \left\{ \sum_{l=1}^{\infty} t_{l}^{(k)} (z_{1}^{l} + \dots + z_{n}^{l}) \right\} \right)
\end{equation}
If we change the sum over $ z_{\alpha} $ in the exponent into product of exponents and, in its turn, the product of exponents into sum over $ t_{m}^{\alpha} $ in the exponent we obtain the following expression
\begin{equation}
    \mathcal{Z}_{n}(\textbf{t}^{(1)}, \dots, \textbf{t}^{(m)}) = \prod_{k=1}^{n} \oint \frac{dz_{k}}{2\pi i} z_{k}^{-i_{k}} \exp \left\{ \sum_{l=1}^{\infty} (t_{l}^{(1)} + \dots + t_{l}^{(m)})z_{k}^{l} \right\},
\end{equation}
where it is possible to calculate each integral. The given exponent is a generating series for symmetric Schur polynomials in variables $ \textbf{t}^{(1)} + \dots + \textbf{t}^{(m)} $ \cite{macdonald1998symmetric}, so contour integral is exactly $ S_{i_{k} - 1} $ for each $ k $ and we obtain the product of $ n $ Schur polynomials:
\begin{equation}\label{generating_function_arbitrary_schurs}
    \mathcal{Z}_{n}(\textbf{t}^{(1)}, \dots, \textbf{t}^{(m)}) = \prod_{k=1}^{n} S_{i_{k} - 1}(\textbf{t}^{(1)} + \dots + \textbf{t}^{(m)}).
\end{equation}
The particular case of $ m=1 $ is the eigenvalue model \eqref{eigenvalue_Schur}. Generating function of the form \eqref{generating_function_arbitrary_schurs} is not very useful to restore coefficients $ \langle S_{\lambda_{1}} \dots S_{\lambda_{n}} \rangle $ since one has to differentiate it with operator $ S(\tilde{\partial}^{(1)}) \dots S(\tilde{\partial}^{(n)}) $ at $ \textbf{t}^{(1)} = \dots = \textbf{t}^{(n)} = 0 $. However, it contains the product of Schur polynomials, which seems similar to the Frobenius formula. The difference between them is that Frobenius formula contains sum over products of Schur polynomials. One may hope that adding times in Schur polynomials as in \eqref{generating_function_arbitrary_schurs} leads to some good properties. 
    
\vspace{0.5cm}
One more question which arises while studying matrix models is the question about any recursion relations. On the one hand we already mentioned recursion relations \eqref{I_recursion} and \eqref{S_recursion}. On the other hand matrix model always has Ward identities, which sometimes can be solved recursively. It turns out that recursion relations obtained from the eigenvalue model are different from both \eqref{I_recursion} and \eqref{S_recursion}. Eigenvalue model \eqref{eigenvalue_model} is provided with Ward identities that give new recursion relations different from \eqref{I_recursion}, \eqref{S_recursion}. We introduce new recursion relations for combinatorial coefficients $P_{i_{1}, \dots, i_{n}}(s_{1}, \dots, s_{m})$ with an arbitrary parameters $ i_{1}, \dots, i_{n} $ and $ s_{1}, \dots, s_{m} $:
\begin{equation}\label{new_recursion_relations}
    P_{i_{1}, \dots, i_{n}}(s_{1}, \dots, s_{m}) = \frac{1}{i_{1} - 1}\sum_{k=1}^{m} \sum_{l=1}^{s_{k}-1} P_{i_{1} - s_{k} + l, i_{2}, \dots, i_{n}}(s_{1}, \dots, s_{k-1}, l, s_{k+1}, \dots, s_{m}).
\end{equation}
As usual for matrix models, Ward identities are obtained with the help of change of variables under the integral that does not change the entire integral. In the case of $ P_{i_{1}, \dots, i_{n}}(s_{1}, \dots, s_{m}) = \langle S_{s_{1}-1} \dots S_{s_{m}-1} \rangle $ in the form \eqref{coefficients_as_average} change of variables is dilatation of the first variable $ z_{1} \rightarrow (1+q)z_{1}, \; q\ne0 $. Integration contour is around $z_{1}=0$, so it is a possible change of variables and deformed integral is independent on $q$. The explicit calculations of deformed integral and proof of \eqref{new_recursion_relations} are presented in Appendix C.

We finish this section with brief review of the obtained results. We provide complete description of combinatorial coefficients $ P_{i_{1}, \dots, i_{n}}(s_{1}, \dots, s_{m}) $ in terms of an eigenvalue model \eqref{eigenvalue_model}. Combinatorial coefficients are certain correlators in the model \eqref{coefficients_as_average} -- averages of symmetric Schur polynomials $ \langle S_{s_{1}-1} \dots S_{s_{m}-1} \rangle $. The full basis of observables is the set of averages of arbitrary Schur polynomials $ \langle S_{\lambda} \rangle $. It is a certain subset of Kostka numbers \eqref{kostka_numbers}. Generating function $ \mathcal{Z}_{n}(\textbf{t}) $ can be calculated explicitly \eqref{generating_function_arbitrary_schurs}. Ward identities give new recursion relations \eqref{new_recursion_relations} for combinatorial coefficients $ P_{i_{1}, \dots, i_{n}}(s_{1}, \dots, s_{m}) $.

\section{Towards deformations of KP hierarchy}

In previous sections we discussed the appearance of coefficients $ P_{i,j}(s_{1}, \dots, s_{m}) $ in the KP equations \eqref{NZ_formulation}. In this section we are interested in a possible generalization of the theory mentioned above, i.e. in the deformation of KP equations. 

As it often happens, various deformations help to understand the underlying structure of the formula, find out which parts are essential and which can be deformed. We try to reveal what role do the combinatorial coefficients $ P_{i,j}(s_{1}, \dots, s_{m}) $ play in equations \eqref{NZ_formulation}. Integrability might be determined by combinatorial coefficients or might be a consequence of the particular form of equations. We deform only the combinatorial coefficients leaving the form of equations unchanged. It turns out that an arbitrary deformation is not possible, equations \eqref{NZ_formulation} contain some restrictions that come from the fact that some of the equations should be fulfilled trivially. These restrictions appear even before the question about compatibility of obtained system of differential equations. However, there is a hopeful deformation direction.

The idea of deformation is based on the fact that we know explicit expression for the generating function of coefficients $ P_{i,j}(s_{1}, \dots, s_{m}) $ of the form \eqref{P_generating_function}. Let us deform this generating function. Deformed coefficients $ P_{i,j}^{(def)}(s_{1}, \dots, s_{m}) $ are obtained as coefficients in the expansion of the deformed generating function similarly to the original ones. At first glance, deformation of the generating function can be done in many ways. For example, we know that generating function \eqref{P_generating_function} is a KP $\tau$-function. So, one can try to let the new generating function be another KP $\tau$-function of a similar form, i.e. $\tau$-function of hypergeometric type \cite{KHARCHEV_1995, orlov2001hypergeometric}. Another way of deformation of generating function is the replacement of Schur polynomials with some other polynomials, which are considered as deformed Schur polynomials, for example, MacDonald polynomials \cite{macdonald1998symmetric}. These two types of generating functions we consider below.


First of all, let us examine which equations in \eqref{NZ_formulation} are trivial. It is obvious that equations \eqref{NZ_formulation} are symmetric due to permutations $ i \leftrightarrow j $. Therefore, we can consider only ordered pair of indices $ i > j $, or, equivalently, equations are labeled by all Young diagrams of length 2. In the case of $ i = n $ and $ j = 1 $:
\begin{equation}\label{trivial_equations}
    \partial_{n}^{\hbar} \partial_{1} F = \sum_{\substack{s_{1} \ge 1 \\ s_{1} = n}} P_{n,1}(s_{1}) \partial_{1} \partial_{s_{1}}^{\hbar}F + \underbrace{ \sum_{\substack{s_{1}, s_{2} \ge 1\\ s_{1} + s_{2} = n-1}} \frac{(-1)}{2 s_{1} s_{2}} P_{n,1}(s_{1},s_{2}) \partial_{1} \partial_{s_{1}}^{\hbar}F \cdot \partial_{1}\partial_{s_{2}}^{\hbar}F + \text{higher m}}_{=0}
\end{equation}
Since one of the indices is equal to 1, there is no matrix of size $ 2 \times m $ with $ m \ge 2 $ with positive integer elements and sum over row equal to 1. Therefore, for any positive integer $ n $ equation \eqref{trivial_equations} reduces to
\begin{equation}\label{trivial_equations_reduced}
    \partial_{1} \partial_{n}^{\hbar}F = P_{n,1}(n) \partial_{1} \partial_{n}^{\hbar}F
\end{equation}
For any $ n $ it is easy to calculate that $ P_{n,1}(n) = 1 $, thus, equations \eqref{trivial_equations_reduced} are hold trivially.

\vspace{0.5cm}
When we replace coefficients $ P_{i,j}(s_{1}, \dots, s_{m}) $ with the deformed ones $ P_{i,j}^{(def)}(s_{1}, \dots, s_{m}) $, the latter ones are calculated via deformed generating function in the following way. Since $ P_{i,j}(s_{1}, \dots, s_{m}) = [x_{1}^{i-1} x_{2}^{j-1} y_{1}^{s_{1}-1} \dots y_{m}^{s_{m}-1}] G(\textbf{x}, \textbf{y}) $, deformed coefficients are obtained similarly as:
\begin{equation}
    P_{i,j}^{(def)}(s_{1}, \dots, s_{m}) = [x_{1}^{i-1} x_{2}^{j-1} y_{1}^{s_{1}-1} \dots y_{m}^{s_{m}-1}] G^{(def)}(\textbf{x}, \textbf{y})
\end{equation}
The deformed equations in the case of $ i = n, j = 1 $ are of the form
\begin{equation}
    \partial_{1} \partial_{n}^{\hbar}F = P_{n1}^{(def)}(n) \partial_{1} \partial_{s_{1}}^{\hbar}F
\end{equation}
and again should be fulfilled trivially. Thus, we have the condition on the deformed coefficients:
\begin{equation}\label{restriction1}
    P_{n1}^{(def)}(n) = 1, \;\; \forall \; n \in \mathbb{N} \;
    \Leftrightarrow \; [x_{1}^{k}y_{1}^{k}]G^{(def)}(\textbf{x}, \textbf{y}) = 1, \;\; \forall \; k \in \mathbb{N} \cup \{ 0 \}
\end{equation}
This condition we consider as a necessary condition for the deformed generating function.

\subsection{Hurwitz deformation}

Let us consider the generating function for simple Hurwitz numbers as a new generating function for combinatorial coefficients. This generating function is a member of the set of hypergeometric $\tau$-functions \cite{kazarian2015combinatorial} and can be written as:
\begin{equation}
    G^{H}(\textbf{x}, \textbf{y}) = \sum_{\lambda} e^{\frac{u}{2} C_{2}(\lambda)} S_{\lambda}(\textbf{x}) S_{\lambda}(\textbf{y})
\end{equation}
where $ C_{2}(\lambda) $ is an eigenvalue of the second Casimir operator \cite{alexandrov2012integrability} ($ C_{2}(\lambda) = \sum_{i=1}^{\ell(\lambda)} \lambda_{i}(\lambda_{i} - 2i + 1) $). The first few terms of the generating function are
\begin{equation}
    G^{H}(\textbf{x}, \textbf{y}) = 1 + (x_{1} + x_{2})(y_{1} + y_{2}) + e^{u}(x_{1}^{2} + x_{1}x_{2} + x_{2}^{2})(y_{1}^{2} + y_{1}y_{2} + y_{2}^{2}) + e^{-u}(x_{1}x_{2})(y_{1}y_{2}) + \dots
\end{equation}
Already in the second order $ [x_{1}^{2}y_{1}^{2}] G^{H}(\textbf{x}, \textbf{y}) = e^{u} $, thus, such deformed coefficients violate necessary condition \eqref{restriction1}. We conclude that Hurwitz numbers is a \textit{bad choice} for deformed combinatorial coefficients. 

\subsection{MacDonald $(q,t)$-deformation}
Although smart $ (q,t) $-deformation of KP hierarchy that possesses an underlying structure of some algebra and solutions like $ (q,t) $-deformed matrix models \cite{Lodin_2019, Cassia_2019} is still unknown, we make an attempt to construct $ (q,t) $-deformed KP equations. Let us consider the sum over MacDonald polynomials as the deformed generating function for combinatorial coefficients:
\begin{equation}\label{MacDonald_sum}
    G^{(q,t)}(\textbf{x}, \textbf{y}) = \sum_{\lambda} M_{\lambda}(\textbf{x}) M_{\lambda}(\textbf{y}),
\end{equation}
Necessary condition \eqref{restriction1} is fulfilled at least for the first few polynomials ($ P_{n1}^{(q,t)}(n) = 1 $ for n = 2,3,4,5), so it is possible that it holds for an arbitrary $n$.

The first non-trivial equation of $(q,t)$-deformed hierarchy is ($ i=2, j=2 $):
\begin{equation}\label{first_qt_equation}
    \partial_{2}^{\hbar} \partial_{2}^{\hbar} F = \frac{4}{3} \left( 1 + \frac{q-t}{1 - qt} \right) \partial_{1} \partial_{3}^{\hbar}F -2 \left( \partial_{1}^{2}F \right)^{2}
\end{equation}
The second non-trivial equation of $(q,t)$-deformed hierarchy is ($ i=3, j=2 $):
\begin{equation}\label{second_qt_equation}
    \partial_{3}^{\hbar} \partial_{2}^{\hbar} F = \frac{3}{2} \left( 1 + \frac{(q-t)(q+1)}{1 - q^{2}t} \right) \partial_{1} \partial_{4}^{\hbar}F -3 \left( \partial_{1}^{2}F \right) \left( \partial_{1} \partial_{2}^{\hbar}F \right)
\end{equation}
Both equations become equations of classical KP hierarchy in the limit $ q = t $. However the question about compatibility of deformed differential equations is still open and deserves a separate study.

Unfortunately, generating function \eqref{MacDonald_sum} does not satisfy equations \eqref{first_qt_equation} and \eqref{second_qt_equation}. Thus, it cannot be considered as a trivial $ \tau $-function of the deformed hierarchy similarly to \eqref{P_generating_function}, which is a trivial $\tau$-function of non-deformed KP. However, the form of the equations remains the same as classical KP hierarchy: each term contains at least two derivatives. Thus, any linear combination of times $t_{k}$ is a solution of these equations. A possible candidate for the deformed trivial $\tau$-function comes from the modification of Cauchy-Littlewood identity \eqref{cauchy_littlewood_identity} for MacDonald polynomials \cite{macdonald1998symmetric}:
\begin{equation}\label{cauchy_littlewood_macdonald}
    \sum_{\lambda} \frac{C_{\lambda}}{C'_{\lambda}} M_{\lambda}(t_{k}) M_{\lambda}(\overline{t}_{k}) = \exp \left( \sum_{k=1}^{\infty} [\beta]_{q} k t_{k} \overline{t}_{k} \right)
\end{equation}
where 
\begin{equation}
    C_{\lambda} = \prod_{(i,j) \in \lambda} \left[ \beta Arm_{\lambda}(i,j) + Leg_{\lambda}(i,j) + 1 \right]_{q}, \;\;\;\;\;\;\;\; C'_{\lambda} = \prod_{(i,j) \in \lambda} \left[ \beta Arm_{\lambda}(i,j) + Leg_{\lambda}(i,j) + \beta \right]_{q}
\end{equation}
Here $[x]_{q}$ denotes the quantum number, $ t = q^{\beta} $ and $ Arm_{\lambda}(i,j), Leg_{\lambda}(i,j) $ are notations of combinatorial objects such as arms and legs of the Young diagram $\lambda$ (for the detailed description of these objects see, for example, \cite{mironov2012proving}). The $F$-function is a logarithm of \eqref{cauchy_littlewood_macdonald} and is just a linear combination of times $t_{k}$ for fixed parameters $\overline{t}_{k}$. Therefore it satisfies deformed equations \eqref{first_qt_equation}, \eqref{second_qt_equation} and might be a possible candidate for a trivial $\tau$-function.

This approach contains some hopeful directions that will be considered in more details elsewhere. Right now generating function \eqref{MacDonald_sum} seems as a \textit{possible choice} for deformed combinatorial coefficients.

\section{Discussion}
In this paper we presented a combinatorial view on the $\hbar$-KP hierarchy based on Natanzon-Zabrodin approach with universal combinatorial coefficients $P_{ij}(s_1,\dots,s_m)$. We showed that studying of the combinatorial coefficients naturally highlights certain properties of the KP hierarchy:
\begin{itemize}
    \item generating function \eqref{P_generating_function} is the KP $\tau$-function by itself and generating function \eqref{H_2} gives Fay identity \eqref{Fay_identity}. These properties give an idea about possible deformations of KP hierarchy from the combinatorial point of view: we expect that deformation of generating function \eqref{P_generating_function} will lead to some interesting deformations of KP hierarchy.
    \item generating function \eqref{H_2} and form of solutions \eqref{F_function_solution} gives information about conditions on Cauchy-like data that corresponds to genus zero resolvents in topological recursion for $\hbar$-KP solutions. In particular, this may be used as a quick test for putative spectral curves for enumerative problems, known to be KP integrable.
    \item combinatorial coefficients $P_{ij}(s_1,\dots,s_m)$ have complete description in terms of quite simple eigenvalue matrix model \eqref{eigenvalue_model}. This approach allows us to describe non-trivial recursion relation \eqref{new_recursion_relations} on the combinatorial coefficients. This matrix model may be used in studying KP hierarchy in terms of the combinatorial coefficients and it gives new questions about interpretation of corresponding averages in terms of KP hierarchy.
\end{itemize}

The aim of this paper is to demonstrate that combinatorial approach to KP hierarchy is instrumental in giving motivation and insights for further study of emergent properties of KP. Here we list some questions that appear naturally when applying this approach:
\begin{itemize}
    \item The question about combinatorial deformation of KP hierarchy is still open: can we deform combinatorial coefficients in equations \eqref{NZ_formulation} in such a way that we obtain an integrable hierarchy? (Discussed in section 6)
    \item What do coefficients $ \langle S_{\lambda_{1}} \dots S_{\lambda_{n}} \rangle $ mean in terms of combinatorial objects or KP hierarchy? (Discussed in section 5)
    \item It is easy to generalize combinatorial definition of the coefficients replacing matrices by tensors. For example, the number of three-tensors with fixed sums over two of three indices is called a Kronecker coefficient, which has a lot of different applications \cite{ikenmeyer2017vanishing, geloun2017tensor}. It is natural to ask, is there any integrable hierarchy formulated via Kronecker coefficients in the same way as the $\hbar$-KP?
    \item How to write a matrix model for such generalizations and how do Ward identities in this model looks like?
    \item According to \cite{Natanzon_2016} it is possible to recover any formal solution of $\hbar$-KP from Cauchy-like data \eqref{f_coefficients} using higher coefficients $P^{\hbar}_{\lambda}
			\begin{pmatrix}
				s_{1} \dots s_{m} \\
				l_{1} \dots l_{m}
			\end{pmatrix}$. Is there any simple combinatorial description for these coefficients? Are they connected with Kronecker numbers in some way? Or may be there is some matrix model generating these coefficients.
\end{itemize}
We hope to address some, or all, of these intriguing questions in the future.

\section*{Acknowledgements}
This work was funded by the Russian Science Foundation (Grant No.20-71-10073). We are grateful to Sergey Fomin and Anton Zabrodin for very useful discussions and remarks. Our special acknowledgement is to Sergey Natanzon for a formulation of the problem and for inspiring us to work on this project.

\section*{Appendix A. Explicit calculation of $P_{i,j}(s_1,\dots,s_m)$}

We start here from the sum that follows from the definition:
\begin{equation}
    P_{i,j}(s_1,\dots,s_m)=\sum\limits_{\{1\leq i_k|k=1,\dots,m\}}\sum\limits_{\{1\leq j_k|k=1,\dots,m\}}\delta_{i_1+\dots+i_m=i}\delta_{j_1+\dots+j_m=j}\delta_{i_1+j_1=s_1+1}\dots\delta_{i_m+j_m=s_m+1}
\end{equation}
Resolving equations $i_k+j_k=s_k+1$  we obtain:
\begin{equation}
    P_{i,j}(s_1,\dots,s_m)=\delta_{s_1+\dots+s_m+m,i+j}\sum\limits_{\{1\leq i_l\leq s_l|l=1,\dots,m\}}\delta_{i_1+\dots+i_m,i}
\end{equation}
Sum in the r.h.s. has geometric interpretation as the section of $m$-dimensional parallelogram $R_{s_1,\dots,s_m}=\{i_k|1\leq i_k\leq s_k, k=1,\dots,m\}$ by $m-1$-dimensional hyper-plane $i_1+\dots+i_m=i$. 
In order to calculate this sum we use inclusion-exclusion principle for $m$-dimensional "quadrants" $Q_{a_1,\dots,a_m}=\{i_k| a_k\leq i_k, k=1,\dots,m\}$. Contribution from $m$-dimensional parallelogram $R_{s_1,\dots,s_m}$ then expressed as the sum over all "quadrants" with vertices coinciding with vertices of $R_{s_1,\dots,s_m}$:
\begin{equation}\label{R_contribution}
    R^{Cont}_{s_1,\dots,s_m}=\sum\limits_{\{\sigma_k=\{0,1\}|k=1,\dots,m\}}(-1)^{\sigma_1+\dots+\sigma_m}Q^{Cont}_{1+\sigma_1s_1,\dots,1+\sigma_ms_m}
\end{equation}
where set of variables $\sigma_k$ enumerate all vertices.

The next step is to calculate contribution of "quadrant" $Q^{Cont}_{1,\dots,1}$, which is just a number of ordered partitions of $i$:
\begin{equation}
    Q^{Cont}_{1,\dots,1}=\sum\limits_{1\leq i_k}\delta_{i_1+\dots+i_m,i}={i-1\choose m-1}.
\end{equation}
Shifting of "quadrant" $Q_{\dots,1,\dots}\rightarrow Q_{\dots,1+s_k,\dots}$ is equivalent to shifting $i\rightarrow i-s_1$, so for the contribution of $Q_{1+\sigma_1 s_1,\dots,1+\sigma_m s_m}$ we have the following formula:
\begin{equation}\label{Q_contribution}
    Q_{1+\sigma_1 s_1,\dots,1+\sigma_m s_m}={i-\sigma_1 s_1 - \dots - \sigma_m s_m -1\choose m-1}
\end{equation}
Combining now \eqref{Q_contribution} and \eqref{R_contribution} we obtain:
\begin{equation}
    R^{Cont}_{s_1,\dots,s_m}  =\sum\limits_{\{\sigma_k=\{0,1\}|k=1,\dots,m\}}(-1)^{\sigma_1+\dots+\sigma_m}{i-\sigma_1 s_1-\dots-\sigma_m s_m-1\choose m-1}
\end{equation}

\section*{Appendix B. Calculation of generating functions}
We give here an approach to calculation of generating functions.

In order to obtain the $G$-generating function \eqref{P_generating_function} it is convenient to use recursion relation \eqref{I_recursion}. Let us substitute \eqref{I_recursion} into the generating function:
\begin{equation}
    \widetilde{G}_{nm}(\mathbf{x},\mathbf{y})=\sum\limits_{i_1\geq 1,\dots,i_n\geq 1}y_1^{i_1}\dots y_n^{i_n}\sum\limits_{s_1\geq 1,\dots, s_m\geq 1}x_1^{s_1}\dots x_m^{s_m}
\sum\limits_{\left\{{i_n^1+\dots+i_n^m=i_n \atop 1\leq i_n^l\leq s_l|l=1,\dots,m}\right\}}P_{i_1\dots i_{n-1}}(s_1-i_n^1+1,\dots,s_m-i_n^m+1)
\end{equation}
The next step is to swap two sums on the right and rewrite each $x_l^{s_l}$ as $x^{i_l^n-1}x_l^{s_l-i_l^n+1}$:
\begin{equation}
    \widetilde{G}_{nm}(\mathbf{x},\mathbf{y})=\sum\limits_{i_1\geq 1,\dots,i_n\geq 1}y_1^{i_1}\dots y_n^{i_n}\sum\limits_{\left\{{i_n^1+\dots+i_n^m=i_n \atop 1\leq i_n^l|l=1,\dots,m}\right\}}x_1^{i_n^1-1}\dots x_m^{i_n^m-1}\sum\limits_{i_n^l\leq s_l\atop l=1,\dots,m}x_1^{s_1-i_n^1+1}\dots x_m^{s_m-i_n^m+1}P_{i_1\dots i_{n-1}}(s_1-i_n^1+1,\dots,s_m-i_n^m+1)
\end{equation}
After replacement $s'_l=s_l-i_n^l+1$ for $l=1,\dots,m$ we obtain simple recursion relation:
\begin{equation}
    \widetilde{G}_{nm}(\mathbf{x},\mathbf{y})=\sum\limits_{i_k\geq 1 \atop k=1,\dot,m}y_1^{i_1}\dots y_n^{i_n}\sum\limits_{\left\{{i_n^1+\dots+i_n^m=i_n \atop 1\leq i_n^l|l=1,\dots,m}\right\}}x_1^{i_n^1-1}\dots x_m^{i_n^m-1}F_{n-1}(\mathbf{x},\mathbf{y})=\widetilde{G}_{(n-1)m}(\mathbf{x},\mathbf{y})\prod\limits_{l=1}^m \frac{y_n}{(1-x_l y_n)}
\end{equation}
where sums over $i_k$ are independent and each of them is geometric progression. It is easy now to write the entire generating function.
\begin{equation}
    \widetilde{G}_{nm}(\mathbf{x},\mathbf{y})=\widetilde{G}_{1m}(\mathbf{x},\mathbf{y})\prod\limits_{l=1}^m\prod\limits_{k=2}^n\frac{y_k}{(1-x_l y_k)},
\end{equation}
where according to our definition of coefficients:
\begin{equation}
    P_{i_1}(s_1,\dots,s_m)=\delta_{s_1+\dots+s_m,i_1}
\end{equation}
and hence
\begin{equation}
    \widetilde{G}_{1m}(\mathbf{x},\mathbf{y})=\sum\limits_{i_1\geq 1}y_1^{i_1}\sum\limits_{s_1\geq 1,\dots, s_m\geq 1}x_1^{s_1}\dots x_m^{s_m}\delta_{s_1+\dots+s_m,i_1}=\prod\limits_{l=1}^m \frac{x_l y_1}{(1-x_l y_1)}.
\end{equation}
Finally, the generating function is of the form:
\begin{equation}
    \widetilde{G}_{nm}(\mathbf{x},\mathbf{y})=\prod\limits_{l=1}^m x_l\prod\limits_{k=1}^n\frac{y_k}{(1-x_ly_k)}=\left(\prod\limits_{l=1}^m x_l\right)\left(\prod\limits_{k=1}^n  y_k^m\right) \sum\limits_{\lambda}S_{\lambda}(\mathbf{x})S_{\lambda}(\mathbf{y})
\end{equation}

Now, using this result we can calculate the second generating function \eqref{H_2}. The main idea is to make replacement $p_k=\sum_{i}x_i^k$:

\begin{equation}
    H(\mathbf{p};y_1,y_2)=\sum\limits_{m\geq 0}\frac{(-1)^{m+1}}{m}\sum\limits_{ij}y_{1}^{i}y_{2}^{j}\sum\limits_{s_1,\dots,s_m}\left(\sum\limits_{i_1}x_{i_1}^{s_1}\right)\dots \left(\sum\limits_{i_m}x_{i_m}^{s_m}\right)P_{ij}(s_1,\dots,s_m)
\end{equation}
Using generating function $\tilde{G}_{2m}$ we obtain
\begin{equation}
\sum\limits_{m\geq 0}\frac{(-1)^{m+1}}{m}\left(\sum\limits_{i_1}x_{i_1}^{s_1}\right)\dots \left(\sum\limits_{i_m}x_{i_m}^{s_m}\right)\prod\limits_{l=1}^m\left(\frac{x_l y_1 y_2}{(1-y_1 x_l)(1-y_2 x_l)}\right).   
\end{equation}
It can be rewritten as the product
\begin{equation}
    \sum\limits_{m\geq 0}\frac{(-1)^{m+1}}{m}\left(\sum\limits_{l=1}^m\frac{x_l y_1 y_2}{(1-y_1 x_l)(1-y_2 x_l)}\right)^m=\left(\frac{y_1 y_2}{y_1-y_2}\sum\limits_{l=1}^m\left(\frac{1}{1-y_1 x_l}-\frac{1}{1-y_2 x_l}\right)\right)^m
\end{equation}
and expanding geometric progression we obtain function in $p_i$ variables:
\begin{equation}
    \sum\limits_{m\geq 0}\frac{(-1)^{m+1}}{m}\left(\frac{y_1 y_2}{y_1-y_2}\sum\limits_{k=1}^{\infty}\left(y_1^k p_k-y_2^k p_k\right)\right)^m=\left(y_1 y_2\sum\limits_{k=1}^{\infty}p_k\frac{y_1^k-y_2^k}{y_1-y_2}\right)^m
\end{equation}
Now summing over $m$ we obtain the generating function.

\bigskip

\section*{Appendix C. Eigenvalue model calculations}
Firstly, we show that expression in brackets in \eqref{almost_average_of_Schurs} is a Schur polynomial. Thus, we prove formula \eqref{coefficients_as_average}. It is obvious that it can be calculated for each $j$ independently, so we do not write index $j$ in the proof. The expression in brackets is equal to
\begin{equation}
    \begin{gathered}
    \frac{z_{n}^{s+n-2}}{z_{1} \dots z_{n-1}} \sum_{i^{(1)} = 1}^{s} \dots \sum_{i^{(n-2)} = 1}^{s} \sum_{i^{(n-1)} = 1}^{s+n-2 - i^{(1)} - \dots -i^{(n-2)}} \left( \frac{z_{1}}{z_{n}} \right)^{i^{(1)}} \dots \left( \frac{z_{n-1}}{z_{n}} \right)^{i^{(n-1)}} \equiv A_{s-1}
    \end{gathered}
\end{equation}
Let us denote the expression as $ A_{s-1} $ and calculate its generating series
\begin{equation}
    A(\xi) = \sum_{s=1}^{\infty} A_{s-1} \xi^{s-1}.
\end{equation}
To perform the calculation we need to swap sum over $ s $ with the other $ (n-1) $ sums over $ i^{(k)} $. All the possible values of indices are inside an $n$-dimensional semi-infinite triangle, and, as usual, changing the order of sums changes the order in which we move inside this triangle with new restrictions on the indices. After swapping the sums one obtains the following expression
\begin{equation}
    A(\xi) = \sum_{i^{(1)} = 1}^{\infty} \dots \sum_{i^{(n-1)} = 1}^{\infty} \sum_{s=i^{(1)} + \dots + i^{(n-1)} - n +2}^{\infty} \frac{z_{n}^{s+n-2}}{z_{1} \dots z_{n-1}} \left( \frac{z_{1}}{z_{n}} \right)^{i^{(1)}} \dots \left( \frac{z_{n-1}}{z_{n}} \right)^{i^{(n-1)}} \xi^{s-1},
\end{equation}
which is now easy to calculate. One has to calculate infinite geometric progressions:
\begin{equation}\label{calculation_a}
    \begin{gathered}
    A(\xi) = \sum_{i^{(1)} = 1}^{\infty} \left( \frac{z_{1}}{z_{n}} \right)^{i^{(1)}} \dots \sum_{i^{(n-1)} = 1}^{\infty} \left( \frac{z_{n-1}}{z_{n}} \right)^{i^{(n-1)}} \cdot \frac{z_{n}^{n-2}}{z_{1} \dots z_{n-1}} \frac{(z_{n} \xi)^{i^{(1)} + \dots + i^{(n-1)} -n + 2}}{\xi (1 - \xi z_{n})} = \\
    = \left( \sum_{i^{(1)} = 1}^{\infty} \frac{1}{z_{1} \xi} (z_{1} \xi)^{i^{(1)}} \right) \dots \left( \sum_{i^{(n-1)} = 1}^{\infty} \frac{1}{z_{n-1} \xi} (z_{n-1} \xi)^{i^{(n-1)}} \right) \cdot \frac{1}{1 - \xi z_{n}} = \prod_{\alpha = 1}^{n} \frac{1}{1 - \xi z_{\alpha}}
    \end{gathered}
\end{equation}
The last expression in \eqref{calculation_a} is exactly a generating function for symmetric Schur polynomials \cite{macdonald1998symmetric}, thus, each $ A_{s-1} $ is equal to Schur polynomial $ S_{s-1} $, which proves \eqref{coefficients_as_average}.

\vspace{0.5cm}
Secondly, we explicitly make clear the derivation of recursion relations \eqref{new_recursion_relations} using the same technique as for Ward identities in common matrix models. Let us rescale the first variable under the integral $ z_{1} \rightarrow (1+q)z_{1} $. There is no singularities except point $ z_{1} = \dots = z_{n} = 0 $, so such a change of variables preserves the value of the integral:
\begin{equation}
    I(q) = \frac{1}{(2 \pi i)^{n}} \oint (1+q)dz_{1} \dots \oint dz_{n} (1+q)^{-i_{1}} \left( \prod_{k=1}^{n} z_{k}^{-i_{k}} \right) \left(  \prod_{j=1}^{m} S_{s_{j}-1}((1+q)z_{1}, \dots, z_{n}) \right)
\end{equation}
This expression is independent on $ q $, so the derivative is equal to zero $ \frac{\partial I}{\partial q} = 0 $. We calculate the derivative at the point $ q = 0 $. Derivative acts on each Schur polynomial independently, so let us first calculate the derivative for only one Schur polynomial:
\begin{equation}\label{second_step}
    \begin{gathered}
    \frac{\partial I(q)}{\partial q} \Bigg|_{q=0} = (1 - i_{1}) P_{i_{1}, \dots, i_{n}}(s) + \oint dz_{1} \dots \oint dz_{n} \left( \prod_{k=1}^{n} z_{k}^{-i_{k}} \right) \left( \frac{\partial}{\partial q} S_{s-1}((1+q)z_{1}, \dots, z_{n}, 0, \dots)\Bigg|_{q=0}  \right) = 0.
    \end{gathered}
\end{equation}
To calculate the derivative of Schur polynomial we use the generating function $A(\xi)$ as in \eqref{calculation_a}, where we rescale the first variable:
\begin{equation}\label{rescaled_generating_function_Schur_polynomials}
    A(q, \xi) = \sum_{j=0}^{\infty} S_{j}((1+q)z_{1}, \dots, z_{n}) \xi^{j} = \frac{1}{1 - (1+q)z_{1}\xi } \prod_{\alpha=2}^{n} \frac{1}{1 - z_{\alpha}\xi}.
\end{equation}
Now it is possible to calculate the derivative of the obtained expression.
\begin{equation}
    \frac{\partial A(q,\xi)}{\partial q} \Bigg|_{q=0} = \frac{z_{1} \xi}{1 - z_{1} \xi} \left( \prod_{\alpha=1}^{n} \frac{1}{1 - z_{\alpha}\xi} \right) = \sum_{j=0}^{\infty} \sum_{p=0}^{\infty} S_{j} z_{1}^{p+1} \xi^{j + p + 1}
\end{equation}
Let us change the summation indices in the last expression: $ a = j+p+1 $.
\begin{equation}
    \frac{\partial A(q,\xi)}{\partial q} \Bigg|_{q=0} = \sum_{a=1}^{\infty} \xi^{a} \left( \sum_{p=0}^{a-1} S_{p} z_{1}^{a-p} \right)
\end{equation}
If we compare it with the expression \eqref{rescaled_generating_function_Schur_polynomials}, we obtain the following result
\begin{equation}
    \frac{\partial S_{a}}{\partial q} \Bigg|_{q=0} = \sum_{p=0}^{a-1} S_{p} z_{1}^{a-p},
\end{equation}
which we substitute into formula \eqref{second_step}:
\begin{equation}
    (1 - i_{1}) P_{i_{1}, \dots, i_{n}}(s) + \oint dz_{1} \dots \oint dz_{n} \left( \prod_{k=1}^{n} z_{k}^{-i_{k}} \right) \left( \sum_{p=0}^{s-2} S_{p} z_{1}^{a-p}  \right) = 0.
\end{equation}
Let us simplify the last expression so it can be rewritten only through combinatorial coefficients:
\begin{equation}
    \begin{gathered}
    0 = (1 - i_{1}) P_{i_{1}, \dots, i_{n}}(s) + \sum_{p=0}^{s-2} \oint dz_{1} \dots \oint dz_{n} z_{1}^{-i_{1}+s-1-p} \left( \prod_{k=2}^{n} z_{k}^{-i_{k}} \right) S_{p} = \\
    = (1 - i_{1}) P_{i_{1}, \dots, i_{n}}(s) + \sum_{p=1}^{s-1} P_{i_{1} - s + p, i_{2}, \dots, i_{n}}(p)
    \end{gathered}
\end{equation}
Now it is easy to do the same calculations for many parameters $ s_{1}, \dots, s_{m} $. Derivative acts on each Schur polynomial labeled by these parameters independently. The result is formula \eqref{new_recursion_relations}.

\bibliographystyle{ieeetr}
\bibliography{references}

\end{document}